=== **INFORMATION TECHNOLOGY IN ENGINEERING SYSTEMS** ===

# Course on System Design (structural approach) ("Information Processes", 2010, 10(4), 303-324)

**Mark Sh. Levin**

*Inst. for Information Transmission Problems, Russian Academy of Sciences,
19 Bolshoj Karetny lane, Moscow 127994, Russia
email: mslevin@acm.org*
Received November 20, 2010

**Abstract**—The article describes a course on system design (structural approach) which involves the following: issues of systems engineering; structural models; basic technological problems (structural system modeling, modular design, evaluation/comparison, revelation of bottlenecks, improvement/upgrade, multistage design, modeling of system evolution); solving methods (optimization, combinatorial optimization, multicriteria decision making); design frameworks; and applications. The course contains lectures and a set of special laboratory works. The laboratory works consist in designing and implementing a set of programs to solve multicriteria problems (ranking/selection, multiple choice problem, clustering, assignment). The programs above are used to solve some standard problems (e.g., hierarchical design of a student plan, design of a marketing strategy). Concurrently, each student can examine a unique applied problem from his/her applied domain(s) (e.g., telemetric system, GSM network, integrated security system, testing of microprocessor systems, wireless sensor, corporative communication network, network topology). Mainly, the course is targeted to developing the student skills in modular analysis and design of various multidisciplinary composite systems (e.g., software, electronic devices, information, computers, communications). The course was implemented in Moscow Institute of Physics and Technology (State University).

### 1. INTRODUCTION

In recent two decades, the following significant trends in engineering have been often considered: (1) systems engineering ([8], [14], [24], [31], [46], [93]); (2) modularity ([6], [28], [37], [40], [83], [101]); (3) issues of system architecture and conceptual design ([9], [24], [38], [40], [101]); (4) usage of decision making, optimization, problem solving and AI techniques ([1], [3], [7], [23], [40], [82], [94], [95], [96], [98], [99]); (5) usage of evolutionary programming (e.g., [105]). The article address an integrated course on system design that is targeted to design some multidisciplinary systems while taking into account the above-mentioned trends.

The described composite educational approach has been implemented as the author course on system design at Faculty of Radio engineering and Cybernetics of Moscow Institute of Physics and Technology (State University) - MIPT ([43], [44], [47], [48], [52]). The above-mentioned university has its well-known experience in the scientific education (e.g., mathematics, physics, information technology) during the first several years of educational process. The goal of the course was to integrate basic knowledge (mathematics, physics, programming) and design approaches (solving schemes/frameworks) into new skills in the study and design of real-world multidisciplinary applied systems/processes. The course was suggested for the fourth year of the educational process (the last year of the undergraduate education). Thus the course is significant for students to conduct their R&D works and to prepare their bachelor theses. It is reasonable to point out Faculty of Radio engineering and Cybernetics at MIPT has a special set of base departments at the research institutes which are engaged in the study and design of applied multidisciplinary systems (e.g., radio engineering, radio physics, computer engineering, applied metrology, VLSI engineering, optical engineering, information processing, communication, control, organizational design, management, socio-economical systems). The composite applied systems above involve various components (software, computers, electronic devices, mechanics, control, human-computer interaction, etc.). Thus various components from different domains have to be examined. Concurrently, many life cycle issues for the systems (R&D, testing, manufacturing, preparation of personnel, marketing, maintenance, utilization, recycling) have to be considered as well.

Note there exist two basic kinds of professional styles (e.g., [26]): (1) *scientific style* (orientation to study and modeling of some applied problems/situations, revelation of some new effects); (2) *engineering style* (orientation to design a goal equipment/system). Contemporary needs in engineering consist in integration of the styles above.

The educational process in MIPT is a very good and unique basis for the realization of the integrated *scientific & engineering style*. Fig. 1 depicts our educational framework that is targeted to the integrated style above.

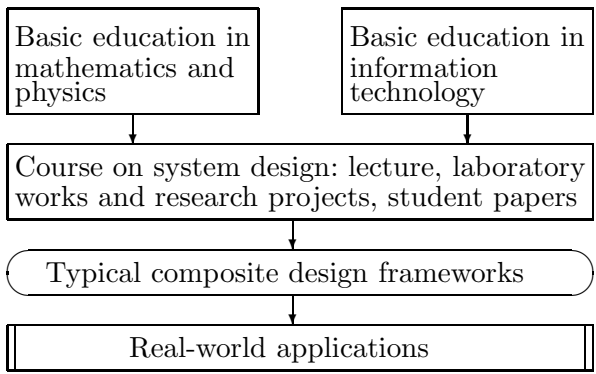

Fig. 1. Generalized educational framework

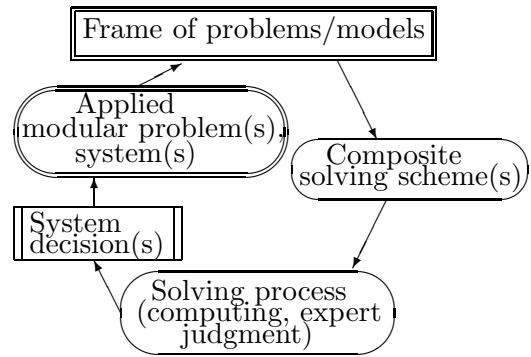

Fig. 2. Decision cycle

At the same time it is necessary to point out, that after the first three years of the basic educational process in MIPT students have often obtained only the basic *scientific style* and movement to *engineering style* is a very hard educational problem. In addition, there are the following educational problems: (i) interdisciplinarity/multidisciplinarity background, (ii) system thinking, (iii) inclination for taking into account system evolution/development and system life cycle, and (iv) inclination for system integration of various components from different applied domains.

In the suggested educational approach, students have to get their experience at the following levels (bottom-up): *1.* modules: decision making and optimization problems and corresponding algorithms/procedures; *2.* solving schemes as design frameworks and composite solving schemes based on modules above; *3.* realistic applied problems. Thus the considered approach is a kind of general (interdisciplinary) engineering education (e.g., [18], [29], [46]) and project-based design engineering education (e.g., [16], [29], [33]). An important feature of the described course consists in information technology components (models, algorithms and procedures, programming environment, applications, Internet). Fig. 2 depicts 'decision cycle' that is used as a basis for our educational efforts. The general design flowchart (composite design scheme) (Fig. 3) is based on the above-mentioned 'decision cycle'.

The examined educational course covers the scheme above (Fig. 3) and contains the following issues ([44], [46]): (i) systems/life cycle engineering; (ii) basic structural system models (structures, graphs, networks, hierarchies, And-Or trees, etc.); (iii) solving schemes (decision making procedures, optimization models and algorithms, methods of artificial intelligence, and design frameworks); (vii) realistic applied examples and real world applications. The layered structure of the course is illustrated in Fig. 4: structural models, solving approaches (optimization, decision making, AI techniques), systems issues, technological problems, and applications [46]. Note a list of the basic technological system problems consists of the following ([46],[50]): (i) modeling, (ii) design/synthesis, (iii) evaluation, (iv) revelation of bottlenecks, (v) redesign (upgrade), (vi) multistage design (or design of a system trajectory), (vii) modeling of system evolution/development (issues of system generations), and (viii) forecasting.

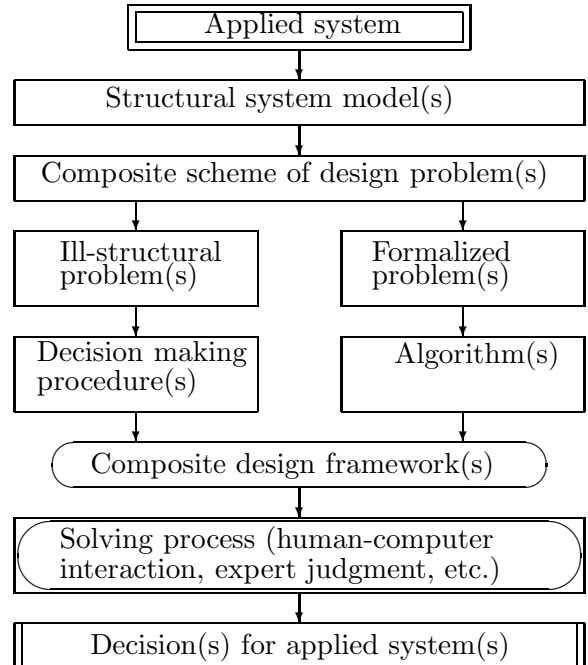

Fig. 3. Composite design scheme

Generally, the considered course corresponds to authors books ([40], [46]) and articles (e.g., [39], [41], [42], [45], [47], [50], [48], [57], [58], [59]). In recent years, a set of student research projects have been published as articles (e.g., [60], [61], [62], [63] [64], [65], [67], [68], [69], [71], [72], [73], [74]). Clearly, basic books in several corresponding domains are used as well, for example: (i) design frameworks (e.g., [3], [5], [9], [15], [32], [101], [107]); (ii) systems engineering (e.g., [31], [37], [92]); (iii) combinatorial optimization (e.g., [4], [11], [22], [25], [35], [84], [85], [90]); and (iv) multicriteria decision making (e.g., [19], [34], [89], [91], [99]).

The course was realized for very good educated and motivated students, but the approach may be used (e.g., as a simplified version) for other students groups as well.

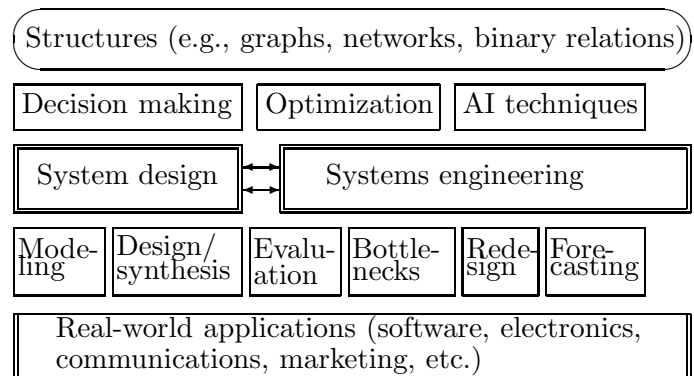

Fig. 4. Layered framework of course

## 2. ORGANIZATION OF COURSE

### 2.1. Course Components

The course includes the following: (1) lectures (about 16...30); (2) simple assignments: design of system model(s), design of Internet personal pages (for each student), and preparation of reports; (3) laboratory works (11 basic works and one additional individual work); (4) special student projects (additional laboratory work): (a) special real world application prepared on the basis of either a laboratory work or modification of a basic algorithm/model, (b) additional work as a special project (additional algorithm/model or/and real-world application); and (5) examination.

The Internet site of the course involves organizational information, lecture notes, bibliography, references to Internet sources (research groups, repositories on problems and algorithms), student lists, references to some student Internet sites [44]. Note the structure of the considered course materials (course ware) is similar to course structures in [80].

### 2.2. Lectures

The lecture plan consists of six parts as follows:

*Part 1.* Basic system issues:

1.1. systems, life cycle, systems engineering, modularity, concurrent engineering;

1.2. structural models as graphs, binary relations; and

1.3. information technology and design frameworks.

*Part 2.* System analysis and decision making:

2.1. principles of system analysis, paradigm of decision making, basic decision making problems, scales, requirements and criteria systems; and

2.2. methods of multicriteria ranking: utility functions, Pareto-approach, outranking techniques.

*Part 3.* Combinatorial optimization and optimization:

3.1. basic combinatorial optimization problems: knapsack, multiple choice knapsack problem, bin-packing and packing, graph coloring, clique, covering, satisfiability, 3-satisfiability, TSP, Hamiltonian cycle, scheduling problems, spanning tree, minimal Steiner tree, alignment, maximal substructure, minimal superstructure, string matching, allocation like problems (assignment problem, quadratic assignment problem, generalized assignment problem, matching problems);

3.2. complexity of combinatorial optimization problems: polynomial solvable problems, NP-complete and NP-hard problems, approximate solvable problems, types of algorithms, methods as global methods and local techniques, heuristics and approximate algorithms, genetic algorithms and evolutionary multiobjective optimization; and

3.3. general optimization model, convex programming, ellipsoid method, mixed integer programming.

*Part 4.* Design frameworks:

4.1. design as generation/evaluation/selection;

4.2. parameter space investigation PSI;

4.3. multidisciplinary optimization;

4.4. hierarchical and cascade-like design; and

4.5. revelation of system bottlenecks.

*Part 5.* Morphological design approaches:

5.1. basic methods: morphological analysis, closeness to ideal solution(s), multicriteria evaluation of all feasible solutions, hierarchical design;

5.2. morphological combinatorial synthesis (Hierarchical Morphological Multicriteria Design HMMD); and

5.3. special applications of HMMD: system improvement (upgrade), multistage design and design of life cycle design of multi-product system, and modeling of system evolution.

*Part 6.* Additional system issues: (system testing, system evaluation/diagnosis, system maintenance, and requirements engineering).

Lecture notes involve many applied examples, e.g., hierarchical morphological design of a student business, hierarchical morphological design of a research team, improvement/upgrade of a team, multicriteria analysis and hierarchical design of a personal computer, multistage evolution of electronic devices, morphological design of a series-parallel solving strategy for multicriteria ranking, modular analysis and evaluation of a building, modeling of system generations for a software system. Communication of the lecturer and students are based on the following: (a) lectures, (b) laboratory works, and (c) Internet (Internet site of the course, email service).

### 2.3. Basic models

A solving (design) process (framework) can be based on one model/procedure or can consist of several interconnected models/procedures (Fig. 3). Fig. 5 illustrates that there are three kinds (levels) of the designed system: product/system, system of requirements, and standard(s). In recent decades the significance of requirements and standards has been increased.

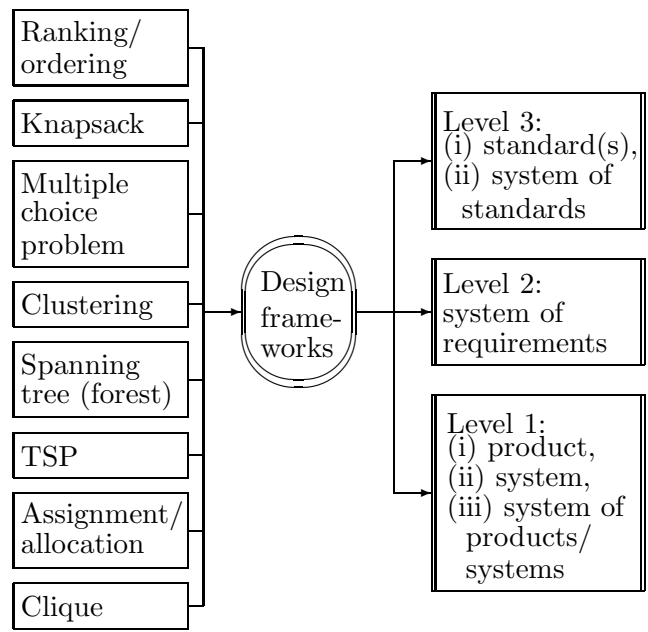

Fig. 5. Basic combinatorial problems, system kinds

### 2.4. Design Approaches

Combinatorial optimization models are used in the course as basic structural procedures for system design. The following three series problems are considered: (1) multicriteria selection/ranking, (2) knapsack problem, and (3) multiple choice problem. The problems above correspond to design framework as selection of design alternatives [15]. Additional problems are assignment and clustering. Assignment/allocation problem corresponds to a design situation when selection of alternatives is integrated with assignemnt/location of the alternatives. Clustering is an important underlying problem for two goals: (i) grouping of elements to decrease problem dimension and (ii) designing a hierarchical structure for a modular system (Bottom-Up method).

Modular morphological approaches are studied for composite (composable) systems. Mainly, HMMD is used for modular combinatorial synthesis (e.g., [39], [40], [43], [44], [46], [51]). HMMD is illustrated for a 3-component system in Fig. 6. Here system $S$ consists of three parts: $A, B, C$. A set of design alternatives (DAs) is examined for each system part: $A_1, A_2, A_3$; $B_1, B_2$; and $C_1, C_2$, Priorities of DAs are shown in the left side (1 corresponds to the best level), quality of compatibility between DAs are shown in the right side (3 corresponds to the best level). Two resultant Pareto-efficient composite DAs are pointed out in Fig. 6: $S_1$ and $S_2$. The following discrete space of the quality for composite systems $S$ is used: $N(S) = (w(S); n_1, n_2, n_3, ...)$, where w(S) corresponds to the worst compatibility in the composite system (between system pats), $n_1$ corresponds to the number of system parts at the best level of quality, etc. The quality of the decisions above are as follows: $N(S_1) = (1; 2, 1, 0)$ and $N(S_2) = (3; 1, 0, 2)$.

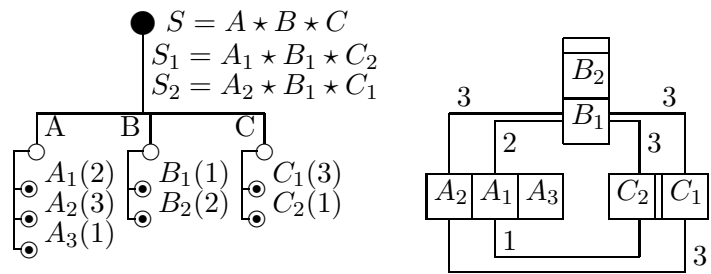

Fig. 6. Example of 3-part system (priorities of DAs are shown in parentheses) [46]

Evidently, the basic system design approaches are described in the course as well, for example: *1.* multidisciplinary optimization MDO (e.g., [2]); *2.* global optimization and mixed-integer non-linear programming ([20], [21]); *3.* parameter space investigation (PSI) ([95], [98]); *4.* morphological analysis ([5], [32], [88]); and *5.* hierarchical decision making ([36], [40]). A special attention is placed to the construction and analysis of applied "engineering design spaces" and corresponding "search spaces" ([12], [40], [100]). This direction may be considered as a crucial one and can help at the stages of problem formulation and solving. Here structural (e.g., hierarchical) models for applied domains, situations, and systems are very important (e.g., [27], [40], [46], [90], [102]).

### 2.5. System Trajectories and Combinatorial Evolution

Multistage system design can be considered as designing a system trajectory. Fig. 7 illustrates three stages of system design. Here a set of resultant decisions is pointed out for each stage (1, 2, and 3):

stage 1 ($t = \tau_1$): $S_1^1, S_2^1, S_3^1$;
stage 2 ($t = \tau_2$): $S_1^2, S_2^2, S_3^2, S_4^2$; and
stage 3 ($t = \tau_3$): $S_1^3, S_2^3, S_3^3, S_4^3$.

The trajectory design process consists in an additional design problem at the higher hierarchical level:

*Select the best decision at each stage while taking into account the compatibility estimates between the decisions for different (e.g., neighbor) stages.*

Examples of trajectories are:

$\alpha_1 = < S_1^1 \to S_4^2 \to S_1^3 >,$
$\alpha_2 = < S_1^1 \to S_2^2 \to S_4^3 >,$
$\alpha_3 = < S_3^1 \to S_4^2 \to S_3^3 >.$

Clearly, the obtained system trajectories can be obtained as Pareto-efficient solutions (e.g., usage of HMMD).

Additional educational efforts can consist in the analysis of system changes (evolution, development). In this case, structural hierarchical models of system generations (e.g., hierarchical models, And-Or trees) are examined and system change operations (from a system generation to the next system generation) can be revealed ([40], [46], [106]). In the course materials, several examples of systems evolution are examined (e.g., four generations of a software package, six generations of an electronic device) ([40], [44], [46]). The system change operations (system improvement actions) can be used as basic modules to realize a next step of system development on the basis of combinatorial optimization problems, e.g., multicriteria selection of the best improvement actions under resource constraint ([46], [54]). Two recent applied examples of combinatorial evolution and development for modular systems are presented in ([66], [70]).

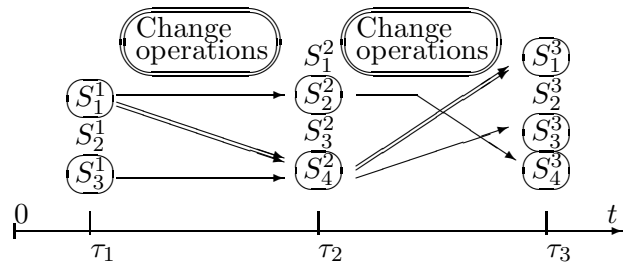

Fig. 7. Illustration of system trajectories

## 3. LABORATORY WORKS

### 3.1. Basic Works

The list of laboratory works involves the following:

Laboratory work 1.

Part 1A: Introductory work (computer environment, design of homepages, preparation of presentations).

Part 1B: Hierarchical morphological design of a system.

Laboratory work 2. Multicriteria ranking (utility function approach, Pareto efficiency approach, method ELECTRE).

Laboratory work 3. Multicriteria knapsack problem.

Laboratory work 4. Method of closeness to ideal point.

Laboratory work 5. Hierarchical clustering.

Laboratory work 6. Multicriteria multiple choice knapsack problem.

Laboratory work 7. Hierarchical ordinal evaluation of composite system (hierarchical method based on integration tables).

Laboratory work 8. Composite applied example: clustering and multiple choice problem (preliminary for work 10).

Laboratory work 9. Assignment/allocation problem.

Laboratory work 10. Composite applied example: clustering & allocation & multiple choice problem.

Laboratory work 11. Travelling salesman problem (TSP).

Laboratory work 12. Additional individual work by choice (e.g., heuristics, genetic algorithms, method based on space filling curves, cross-entropy method, quadratic assignment problem, covering problems, real world applications).

Table 1. Laboratory works and papers

| Laboratory works | Support materials | Students papers |
|---|---|---|
| 1. Design of modular system:<br>  (a) hierarchical system model,<br>  (b) morphological design | [5],[32],[40],<br>[41],[42],[45],<br>[46],[56],[88],<br>[90],[102] | [61],[62],<br>[64],[65],<br>[72] |
| 2. Multicriteria ranking:<br>  (1) utility function approach,<br>  (2) Pareto-efficiency method,<br>  (3) outranking technique<br>    (ELECTRE-like method) | [19],[34]<br>[86]<br>[89] | |
| 3. Multicriteria knapsack problem | [35] | |
| 4. Method of closeness to ideal point | [103],[104] | |
| 5. Hierarchical clustering | [30],[49],[79] | |
| 6. Multicriteria multiple choice knapsack problem | [35],[54],[56],<br>[97] | [60],[68],<br>[71],[73] |
| 7. Hierarchical evaluation of composite system | [46],[58] | |
| 8. Two-problem framework (preliminary for work 10) | | |
| 9. Assignment/allocation problems | [10],[11],[40],<br>[51],[85],[87] | [69] |
| 10. Four-problem framework:<br>  (i) clustering (for two sets),<br>  (ii) assignment,<br>  (iii) multiple choice knapsack<br>    problem | [53] | [63] |
| 11. TSP | [4],[25] | |
| 12. Work by individual choice:<br>  (i) new models;<br>  (ii) new methods/algorithms;<br>  (iii) algorithm(s) analysis;<br>  (iv) new applications | [11],[13],[22],<br>[40],[51],[66],<br>[70] | [67],[68],<br>[74] |

In each laboratory work, students have to do the following: (1) to understand the material (problem, model, solving scheme as algorithm, framework), (2) to develop software and to test it, (3) to prepare a numerical example (or a real world application), (4) to compute the results, and (5) to prepare the report. For some basic laboratory works there are some basic program (in MatLab [78]), for example: multicriteria ranking (utility function method, revelation of Pareto-effective solutions, outranking technique as a modification of method Electre), heuristic for knapsack problem, heuristic for multiple choice problem, hierarchical clustering (agglomerative algorithm). As a result, each student has to improve the basic program to get his/her resultant program. For more complicated (i.e., composite) laboratory works (for example: 1, 6, 8, 10) each student can combine his/her programs to get the resultant composite software. Table 1 contains correspondence between laboratory works and literature sources (including the authors books and articles) and published student articles/papers.

Generally, it is possible to point out several layers of considered problems/methods:

Layer 1. Basic combinatorial problems and multicriteria DM-problems: multicriteria ranking, multicriteria knapsack problem, 'ideal point' method, system evaluation/diagnosis.

Layer 2. Advanced models: (clustering, multicriteria multiple choice problem, multicriteria assignment/allocation, TSP).

Layer 3. Composite models/procedures (as basic solving composite scheme consisting of problems/models): (i) HMMD (ranking, combinatorial synthesis) to design modular systems, (ii) design of a modular solving strategy (e.g., a partitioning/synthesis heuristic for problem), (iii) system up-grade, (iv) multistage design (two-level HMMD), (v) system evolution/forecasting, (vi) special multistage composite framework (clustering, assignment/location, multiple choice knapsack problem).

Fig. 8 depicts a flowchart over the laboratory works.

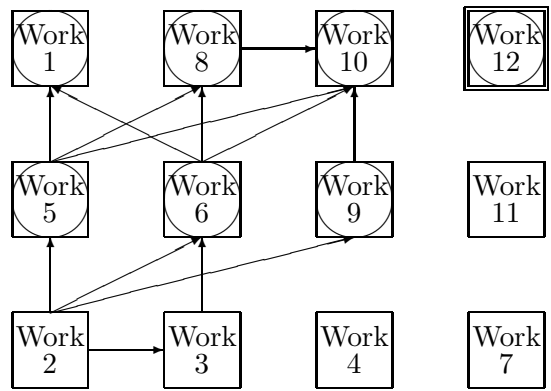

Fig. 8. Flowchart over laboratory works

### 3.2. Composite Works

The basic list of laboratory works involves two composite works (composite frameworks):

*Composite work 1.* Hierarchical modular design of a student strategy (HMMD) (*work 1b*):

*Step 1.* Design of a hierarchical system model.

*Step 2.* Generation of DAs for leaf nodes of the model.

*Step 3.* Generation of criteria sets for each model node.

*Step 4.* Multicriteria ranking of the alternatives (*work 2*).

*Step 5.* Bottom-up synthesis of composite alternatives.

Here a basic illustrative applied example consists in a hierarchical student strategy (Fig. 9) [40]. Each student has to design his/her career plan, consisting of selected alternatives for the following parts: basic courses at different levels (e.g., mathematics, operations research, management), advanced courses, additional courses (e.g., history, languages, psychology), sport activity (e.g., body building, karate, tennis), art activity (e.g., classical music, jazz, dance, theater), and work (a temporary work and a prospective professional work).

*Composite work 2.* Three-set & four-problem strategy (*work 10*) (Fig. 10):

*Step 1.* Clustering the elements of the 1st set (*work 5*).

*Step 2.* Clustering the elements of the 2nd set (*work 5*).

*Step 3.* Assignment/allocation of clusters of the 1st set into clusters of the 2nd set, (assignment problem or generalized assignment problem, *work 9*).

*Step 4.* Selection of an activity (the 3rd set of activities) for each interconnection "element of 1st set - element of 2nd set" (multiple choice knapsack problem, *work 6*).

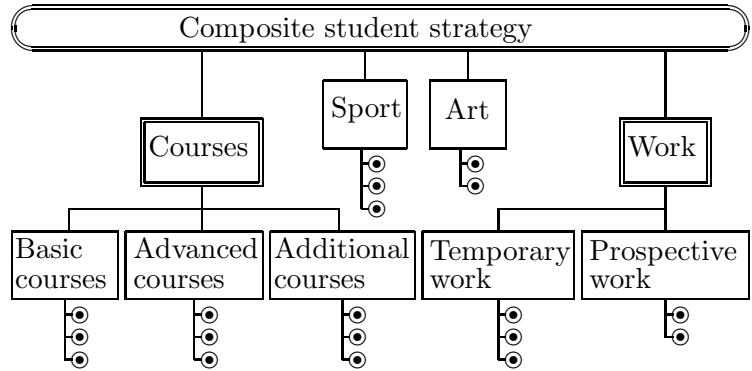

Fig. 9. Hierarchy of student strategy

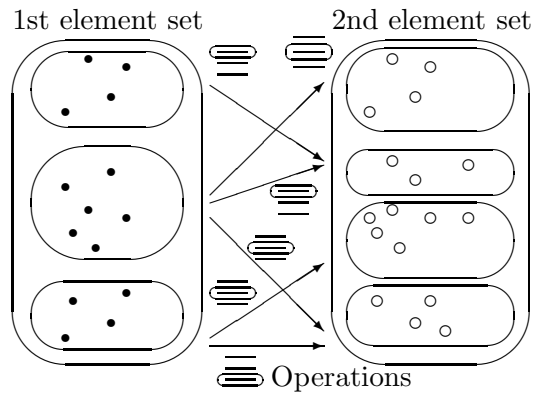

Fig. 10. Composite 3-set 4-problem strategy

### 3.3. Additional Laboratory Work

Basic topics for special student projects (*work 12*) are the following: (i) assignment/ allocation problems and algorithms, (ii) timetabling problems, (iii) analysis and usage of cross-entropy method, (iv) usage of genetic algorithms, (v) analysis and usage of space-filing curve approach, (vi) special applied examples for system technological problems (e.g., system improvement, system evolution, system forecasting), and (vii) special applications based on student experience and/or intention(s). The goals of this work kind are to get skills in the following: (a) searching for useful information, (b) understanding and structuring new information, (c) research, and (d) creating some new models (algorithm, solving schemes, problem formulations). Here the student can choose a level of his/her research: (1) a basic research to study algorithm(s), software, realistic applied example(s), composite solving scheme(s) and/or (2) an advanced research: (a) survey of well-known approaches/models/algorithms, (b) investigation and design of new model(s) or solving scheme.

## 4. COMPUTER ENVIRONMENT

The course is based on computer environment that consists of the following parts:

*1.* Computer class:

    *1.1.* Computer network and software environment, including MatLab environment [78], Internet.

    *1.2.* Programs (support programs, prototype-programs).

    *1.3.* Additional materials: *1.3.1.* text-books, *1.3.2.* articles, *1.3.3.* student materials on models and applications (e.g., reports on laboratory works), *1.3.4.* student theses/materials, *1.3.5.* references (sites, bibliography-sources), *1.3.6.* special applied examples and case studies.

  *2.* Student home computer environment.

  *3.* Internet course site [44], including the following:

    *3.1.* Course description.

    *3.2.* Lecture notes (slides).

    *3.3.* Assignments.

    *3.4.* Additional materials (bibliography, topics, etc.).

    *3.5.* Hypertext on models, technological problems, and applied examples.

    *3.6.* Internet sources (sites on models/algorithms, design topics, etc.).

  *4.* Student Internet home pages.

The site of course [44] (and the site of the author [43]) contains a set of research topics and corresponding references to Internet sites (research groups, repositories of problem descriptions and algorithms, etc.). Some examples of the Internet sites on research topics are the following:

(i) multidisciplinary optimization MDO [77],

(ii) maintenance [76],

(iii) evolutionary multiobjective optimization EMO [17],

(iv) quadratic assignment problem [87], and

(v) cross-entropy method [13].

It can be very important to give students an opportunity to search for and to evaluate useful information. Parts 3.4 and 3.6 above are targeted to the above-mentioned student activity. Designing the student home pages (part 4 above) may be considered as an additional evaluation and design work for the students. The design of the student home page is a crucial process, because each student can define/understand his/her professional and personal profile.

## 5. LABORATORY WORKS AS BASIS FOR PROJECTS

### 5.1. Educational Flowchart for Research

The movement from traditional educational process to research and/or design activity is often the critical part of many educational efforts (computer science, information technology, engineering, applied mathematics). In the article, our approach is described for the above-mentioned movement. Our educational suggestion is based on a set of laboratory work in which an educational flow is used: applied problem, mathematical model, algorithm, software, computing the results, and report (Fig. 11).

In each work a special combinatorial and/or multicriteria problem is considered (e.g., multicriteria ranking, multiple choice problem, assignment/allocation, clustering) including basic applied examples and algorithm(s) (and basic program in Matlab). As a result, each student can understand all aspects of the problem. In addition, each student can take into account his/her inclination, motivations, and personal goals. After that the student can select a part of the educational flow above to prepare a modified or new version of the part (i.e., applied problem, model, algorithm). Concurrently, student can obtain an important skills in composition of problems/models/algorithms to get a composite framework for complicated real world applications.

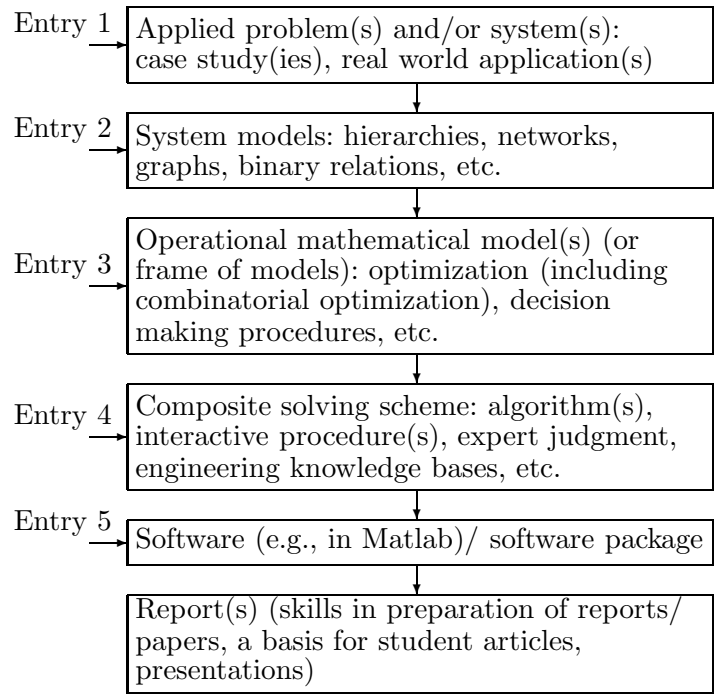

Fig. 11. Flow-chart based on laboratory works

### 5.2. Student Movement to Research

Composite laboratory works are used as a fundamental for student research projects. Generally, there are four possibility for the student research: (a) real world application (an engineering research), (b) new or modified model (research in mathematics or computer science), (c) new or modified algorithm (solving scheme) (research in mathematics or computer science), (d) new or modified program (research in applied computer science). As a result, each student can choose the certain kind of his/her research. This choice process is based on student inclination and/or experience.

In the simplest situation, students can use a simple numerical version of an applied problem (i.e., multicritertia ranking, knapsack, multiple choice problem, assignment, TSP).

On the other hand, each student can examine a special real world application. This step is crucial for composite problems in the following laboratory works: 1, 5, 6, 7, 9, 10, 11, 12. The example list of student research projects involves the following:

(1) sport applications (e.g., organization of sport events, planning of sport activity),

(2) educational application (computer class, educational interactive software),

(3) software development (e.g., software for modeling of signals),

(4) hardware applications (e.g., computer memory),

(5) communication systems (analysis of communication protocols, routing, devices for computer networks, connection of clients to network, improvement or extension of a network),

(6) sensors and wireless systems (telemetry system, wireless sensor),

(7) VLSI design (planning of test processes, planning of manufacturing),

(8) private electronic devices (digital camera, notebooks, cars, mobile phone), and

(9) Web-based application (information systems, Internet-based station), etc.

In the case of student inclination for algorithmic studies, it is possible to design new or modified algorithms. Some students have investigated new models, for example: multicriteria Steiner tree model ([67], [74]). It is reasonable to point out some topics of students research projects which were based on laboratory work 12 (by choice): (a) algorithms for routing in communication networks; (b) experimental comparison of heuristics for TSP or quadratic assignment problem (e.g., genetic algorithm, ant colony optimization, cross entropy method); (c) scheduling in wireless sensor networks.

Finally, the course has several entries for student research projects (Fig. 11):

Entry 1 (application): new real world application (on the basis of existing problem(s), model(s), algorithm(s), software).

Entry 2 (system mathematical models as hierarchy, network, etc.): model(s) (new or modified model).

Entry 3 (operational mathematical models as optimization, etc.): model(s) (new or modified mathematical model).

Entry 4 (algorithms): solving scheme and/or their analysis (new or modified algorithm/solving scheme, theoretical and/or experimental study).

Entry 5 (software): program (new or modified software/software package).

As a result, each student can choose the corresponding entry to his/her research and this choice process is based on student inclination/experience.

### 5.3. Examples of Realistic Applied Student Works

Since September 2004, many students have used design approaches based on laboratory works (e.g., combinatorial problems and composite frameworks) in their applied projects (special applied examples):

*1.* morphological combinatorial synthesis based on HMMD (work 1): telemetric system [62], GSM communication network [61], integrated security system [65], wireless alarm sensor [72], and information Web-hosting system [64];

*2.* multicriteria multiple choice knapsack problem (work 6): improvement of communication network [71];

*3.* assignment/allocation problem (work 9): assignment of users to access points in telecommunication network [69];

*4.* multicriteria Steiner tree problem (work 12): design of network topology ([67], [74]);

*5.* composite four-problem framework (work 10): planning of testing process for microprocessor systems [63].

The above-mentioned student research projects extend the basic course materials.

### 6. BASIC EDUCATIONAL DESIGN PROBLEMS

Fig. 12 illustrates interconnection between basic course materials and design problems (five) for lecturer and student(s). These educational design problems above may be based on HMMD. Examples of two lecturer's design problems are considered in [40]: design of curriculum and design of course environment. Two student personal design problems are included into the course: (i) plan of career, (ii) plan of business ([40], [43]). Student applied design problems are considered under framework of laboratory works and student projects/articles (design of information systems, programs, communication networks, etc.). The design of a student individual research-learning environment is a very prospective educational direction. It can be vital to organize this design process for each student.

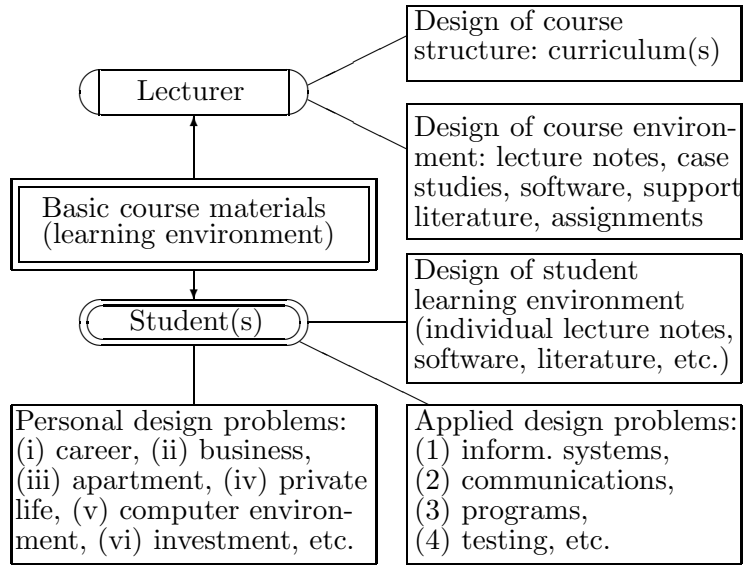

Fig. 12. Design problems for lecturer and for student

## 7. ILLUSTRATIVE EXAMPLES OF COMPOSITE WORKS

The first illustrative example is: designing a version of the course (Fig. 13, laboratory work 1). This example can be useful for lecturer (a course version for a student group ) or for a student (an individual course version).

The initial structure $S = E \star H \star W$ is:

*Part 1.* Systems issues $E = L \star M \star F \star G$: issues of systems engineering $L$, system modularity $M$, design frameworks $F$, and structured models $G$.

*Part 2.* Decision making and optimization $H = D \star O \star B$: methods of multicriteria decision making $D$, optimization $O$, and combinatorial optimization $B$.

*Part 3.* Morphological design $W = P \star I \star C$: hierarchical design $P$, system improvement $I$, and morphological framework for assignment/allocation $C$.

Here four design alternatives are examined for each leaf node: none $X_1$, brief description/explanation $X_2$, explanation and laboratory work $X_3$, explanation and research project $X_4$. Some resultant estimates as ordinal priorities for DAs are depicted in Fig. 13 in parentheses (expert judgment, scale $[1,3]$). Tables 2, 3, and 4 contain estimates of compatibility (scale $[0,3]$, 3 corresponds to the best level of compatibility; expert judgment). Evidently, the ordinal priorities can be obtained by multicriteria ranking of multicriteria estimates of DAs. The same approach can be used to get ordinal compatibility estimates. Composite DAs for system parts are the following:

*Part 1:* $E_1 = L_2 \star M_2 \star F_2 \star G_3$, $N(E_1) = (2; 4, 0, 0)$,
$E_2 = L_3 \star M_3 \star F_2 \star G_3$, $N(E_2) = (3; 2, 2, 0)$.
*Part 2:* $H_1 = D_3 \star O_3 \star B_3$, $H_2 = D_3 \star O_3 \star B_4$, $N(H_1) = N(H_2) = (3; 3, 0, 0)$.
*Part 3:* $W_1 = P_4 \star I_3 \star C_3$, $N(W_1) = (3; 3, 0, 0)$.

The resultant composite DAs for the course are the following (without an analysis of quality of DAs and compatibility):
$S_1 = E_1 \star H_1 \star W_1$, $S_2 = E_1 \star H_2 \star W_1$,
$S_3 = E_2 \star H_1 \star W_1$, and $S_4 = E_2 \star H_2 \star W_1$.

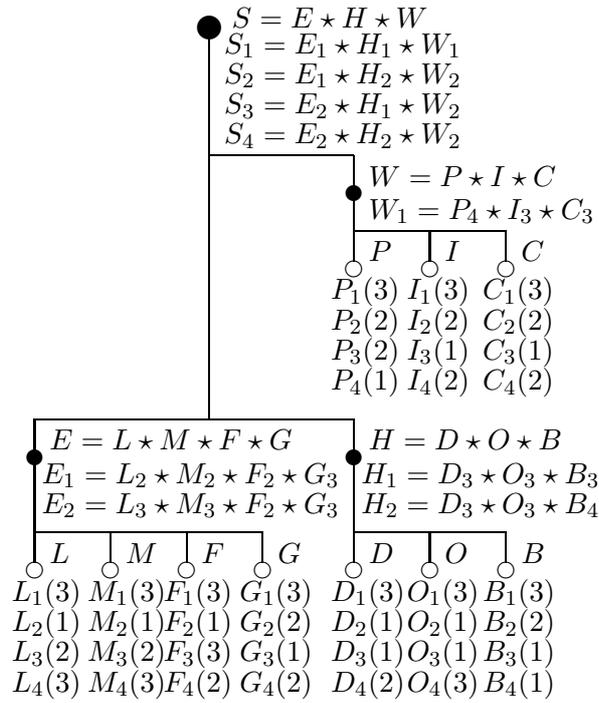

Fig. 13. Structure of course

Table 2. Compatibility estimates

|   | $M_1$ | $M_2$ | $M_3$ | $M_4$ | $F_1$ | $F_2$ | $F_3$ | $F_4$ | $G_1$ | $G_2$ | $G_3$ | $G_4$ |
|---|---|---|---|---|---|---|---|---|---|---|---|---|
| $L_1$ | 1 | 2 | 0 | 0 | 1 | 2 | 0 | 0 | 1 | 1 | 1 | 1 |
| $L_2$ | 1 | 2 | 1 | 1 | 1 | 2 | 2 | 1 | 1 | 1 | 2 | 2 |
| $L_3$ | 0 | 3 | 3 | 3 | 0 | 3 | 3 | 3 | 1 | 2 | 3 | 3 |
| $L_4$ | 0 | 3 | 3 | 3 | 0 | 3 | 3 | 3 | 1 | 2 | 3 | 3 |
| $M_1$ |   |   |   |   | 1 | 2 | 0 | 0 | 1 | 2 | 0 | 0 |
| $M_2$ |   |   |   |   | 1 | 2 | 2 | 0 | 1 | 2 | 3 | 3 |
| $M_3$ |   |   |   |   | 0 | 3 | 3 | 3 | 0 | 3 | 3 | 3 |
| $M_4$ |   |   |   |   | 0 | 3 | 3 | 3 | 0 | 3 | 3 | 3 |
| $F_1$ |   |   |   |   |   |   |   |   | 1 | 1 | 0 | 0 |
| $F_2$ |   |   |   |   |   |   |   |   | 1 | 2 | 3 | 3 |
| $F_3$ |   |   |   |   |   |   |   |   | 0 | 2 | 3 | 3 |
| $F_4$ |   |   |   |   |   |   |   |   | 0 | 2 | 3 | 3 |

Table 3. Compatibility estimates

|   | $O_1$ | $O_2$ | $O_3$ | $O_4$ | $B_1$ | $B_2$ | $B_3$ | $B_4$ |
|---|---|---|---|---|---|---|---|---|
| $D_1$ | 1 | 2 | 0 | 0 | 1 | 1 | 0 | 0 |
| $D_2$ | 2 | 3 | 3 | 3 | 2 | 2 | 2 | 2 |
| $D_3$ | 0 | 3 | 3 | 3 | 2 | 3 | 3 | 3 |
| $D_4$ | 0 | 3 | 3 | 3 | 2 | 3 | 3 | 3 |
| $O_1$ |   |   |   |   | 1 | 1 | 0 | 0 |
| $O_2$ |   |   |   |   | 1 | 2 | 2 | 2 |
| $O_3$ |   |   |   |   | 0 | 2 | 3 | 3 |
| $O_4$ |   |   |   |   | 0 | 2 | 3 | 3 |

Table 4. Compatibility estimates

|   | $I_1$ | $I_2$ | $I_3$ | $I_4$ | $C_1$ | $C_2$ | $C_3$ | $C_4$ |
|---|---|---|---|---|---|---|---|---|
| $P_1$ | 1 | 1 | 0 | 0 | 1 | 2 | 0 | 0 |
| $P_2$ | 1 | 1 | 2 | 2 | 2 | 3 | 3 | 3 |
| $P_3$ | 2 | 3 | 3 | 3 | 0 | 3 | 3 | 3 |
| $P_4$ | 2 | 3 | 3 | 3 | 0 | 3 | 3 | 3 |
| $I_1$ |   |   |   |   | 3 | 3 | 3 | 2 |
| $I_2$ |   |   |   |   | 3 | 3 | 3 | 3 |
| $I_3$ |   |   |   |   | 3 | 3 | 3 | 3 |
| $I_4$ |   |   |   |   | 2 | 3 | 3 | 3 |

The second illustrative example corresponds to a simplified version of laboratory work 10. There are five PhD students ($A_1$, $A_2$, $A_3$, $A_4$, $A_5$) and four laboratory works ($V_1$, $V_2$, $V_3$, $V_4$). It is necessary to select a student for each laboratory work (for role of teaching assistant) and select a level of the teaching process: explanation $T_1$, advising the laboratory work $T_2$, and joint research project $T_3$. At the first stage, the assignment problem has to be solved to obtain pairs: (student, work). A correspondence of student to work consists of three parameters (ordinal scale [0,7]): (a) theoretical level $c^1$, (b) engineering experience $c^2$, and (c) research experience $c^3$.

At the second stage, for each pair (student-work) it is necessary to select the level of teaching while taking into account a total budget (multiple choice knapsack problem). Table 5 contains estimates of the correspondence: $\{(c_{ij}^1, c_{ij}^2, c_{ij}^3)\}$. Clearly, a three criteria assignment problem is obtained. Thus the assignment problem is:

$$\max \sum_{i=1}^{5}\sum_{j=1}^{4} c_{ij}^1 x_{ij}, \quad \max \sum_{i=1}^{5}\sum_{j=1}^{4} c_{ij}^2 x_{ij}, \quad \max \sum_{i=1}^{5}\sum_{j=1}^{4} c_{ij}^3 x_{ij}$$

$$s.t. \sum_{i=1}^{5} x_{ij} \leq 1, j=\overline{1,4}; \ \sum_{j=1}^{4} x_{ij} = 1, i=\overline{1,5}; \ \ x_{ij} \in \{0,1\}.$$

A resultant assignment solution is (by heuristic): $A_1 \to V_6$ (corresponding Boolean variable for multiple choice problem $y_{1j}$), $A_2 \to V_{10}$ ($y_{2j}$), $A_3 \to V_{12}$ ($y_{3j}$), and $A_5 \to V_1$ ($y_{4j}$). In general, here several Pareto-efficient solutions can be obtained.

Table 5. Correspondence: 'student - work'

|       | $V_1$     | $V_6$     | $V_{10}$  | $V_{12}$  |
|-------|-----------|-----------|-----------|-----------|
| $A_1$ | (2,2,1)   | (2,4,6)   | (2,2,2)   | (4,2,3)   |
| $A_2$ | (2,1,1)   | (2,1,2)   | (2,3,7)   | (2,4,2)   |
| $A_3$ | (2,3,1)   | (2,2,2)   | (2,2,2)   | (4,1,6)   |
| $A_4$ | (1,2,1)   | (2,3,2)   | (2,1,2)   | (2,2,6)   |
| $A_5$ | (1,7,2)   | (2,2,1)   | (1,1,1)   | (1,3,1)   |

Let a cost (e.g., money, time; $a_{ij}$) of the teaching activities be the following: 2 for $T_1$, 3 for $T_2$, 4 for $T_3$. The structure of multiple choice problem is depicted in Fig. 14 and the problem is ($b$ is a constraint):

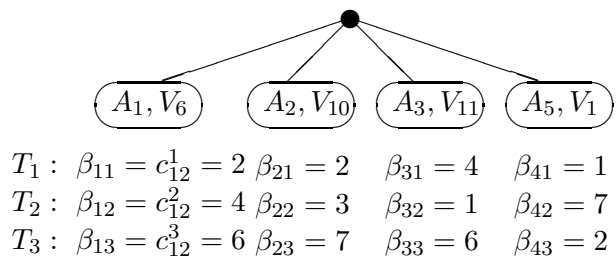

$T_1: \ \beta_{11} = c_{12}^1 = 2 \ \ \beta_{21} = 2 \ \ \beta_{31} = 4 \ \ \beta_{41} = 1$
$T_2: \ \beta_{12} = c_{12}^2 = 4 \ \ \beta_{22} = 3 \ \ \beta_{32} = 1 \ \ \beta_{42} = 7$
$T_3: \ \beta_{13} = c_{12}^3 = 6 \ \ \beta_{23} = 7 \ \ \beta_{33} = 6 \ \ \beta_{43} = 2$

Fig. 14. Structure of multiple choice problem

$$\max \sum_{i=1}^{4}\sum_{j=1}^{3} \beta_{ij} y_{ij}$$

$$s.t. \ \sum_{i=1}^{4}\sum_{j=1}^{3} a_{ij} y_{ij} \leq b, \ \ \sum_{j=1}^{3} y_{ij} \leq 1 \ \forall i = \overline{1,4}, \ \ y_{ij} = \{0,1\}.$$

Some illustrative solutions are (by heuristic):

(i) $b = 15$:

$(A_1, V_6, T_3), (A_2, V_{10}, T_3), (A_3, V_{12}, T_3), (A_5, V_1, T_2)$;
(ii) $b = 12$:
$(A_1, V_6, T_3), (A_2, V_{10}, T_3), (A_3, V_{12}, T_1), (A_5, V_1, T_2)$;
(iii) $b = 10$: $(A_1, V_6, T_2), (A_2, V_{10}, T_3), (A_5, V_1, T_2)$.

## 8. CONCLUSION

The considered course is targeted to integration of two kinds of styles: *scientific* style and *engineering* style. During the course, students can obtain their ability to design/redesign new complex systems. Concurrently, they are obtaining experience in the following: (i) design of structural (e.g., hierarchical) models for systems and applied situations; (ii) ability to design problem solving schemes, including models of search spaces, decision making procedures, optimization framework, etc.; and (iii) preparation of reports and presentations.

Generally, it is reasonable to point out a basic trend "from simple problem(s) to complex problem(s)". This movement in education can be based on examination of $k$-set frameworks [55] (Fig. 6): one-set frameworks, two-set frameworks, three-set frameworks, etc. This trend can allow (step-by-step) a series extension of students experience and skill in formulation and solving of applied complicated problems.

Table 6. $k$-set frameworks and problems

| Type of framework | Problems/works | Basic set(s) |
|---|---|---|
| One-set framework | 1.Ranking (work 2) 2.Knapsack (work 3) 3.Clustering (work 5) | Set of elements (alternatives, items) |
| Two-set framework | Assignment/allocation (work 9) | Set of elements, set of positions |
| Three-set framework | Four-problem framework (work 10) | Set of elements, set of positions, set of actions |

The described course material provides a basis for the analysis and design of contemporary multidisciplinary high-tech applied systems while taking into account issues of life cycle engineering. In the author opinion, a communication skill in multidisciplinary distributed engineering environment is very important. From this viewpoint, it will be reasonable to organize student projects as follows: *1.* multidisciplinary student groups (e.g., students from different departments/faculties: engineering, computer science, mathematics, management); *2.* distributed student groups (e.g., students from different universities).

As an extension of the course materials it can be prospective the following: (i) examination of various applied problems from student life and preparation of publications, for example, student research results on designing in sport; (ii) special attention to design issues at the levels of system requirements and standards; (iii) study and usage of the considered models under uncertainty (i.e., probabilistic models, usage of fuzzy sets); (iv) wide usage of AI techniques; (v) conducting some student research projects on laboratory works 2, 3, 4, 7, 11 to publish student articles on the topics; (vi) analysis of additional multi-set frameworks and their application in real-world problems; (vii) inclusion into course the concept of "systems of systems" (e.g., [31], [75]); and (viii) improvement of the course site.

Finally, it can be very useful to implement the suggested course on system design at the level of graduate studies and continuous education.

# 9. ACKNOWLEDGMENT

The author acknowledges the administration of Faculty of Radio Engineering and Cybernetics at MIPT. The author thanks Academician, Prof. N.A. Kuznetsov (Chair, Dept. of Communication Networks and Systems at MIPT) for his attention. The course was partially supported by NetCracker Inc. (USA) [81].